\documentclass[usenatbib]{mn2e}
\pdfoutput=1
\usepackage[totalwidth=480pt,totalheight=680pt]{geometry}
\usepackage{graphicx}
\usepackage{latexsym}
\usepackage{aas_macros}
\usepackage{amsfonts,amsmath,amssymb}
\usepackage{url}
\usepackage[utf8]{inputenc}
\usepackage{fancyref}

\title[A parameter study of the eclipsing CV in the Kepler field,
KIS J192748.53+444724.5]{A parameter study of the eclipsing CV in the Kepler field,
KIS J192748.53+444724.5}

\author[Stuart Littlefair]{S.\,P.\ Littlefair$^{1}$, V.\,S.\ Dhillon$^{1}$, B.\,T.\ G\"{a}nsicke$^{2}$, M.\,C.\,P.\ Bours$^{2}$, 
\newauthor C.\,M.\ Copperwheat$^{3}$, T.\,R.\, Marsh$^{2}$\\
$^{1}$Dept of Physics and Astronomy, University of Sheffield, Sheffield, S3 7RH, UK \\
$^{2}$Dept of Physics, University of Warwick, Coventry, CV4 7AL, UK\\
$^{3}$Astrophysics Research Institute, Liverpool John Moores University, Twelve Quays House, Birkenhead CH41 1LD, UK}

\begin{document}



\maketitle

\label{firstpage}

\begin{abstract}
We present high-speed, three-colour photometry of the eclipsing
dwarf nova KIS J192748.53+444724.5 which is located in the
Kepler field. Our data reveal sharp features corresponding to the
eclipses of the accreting white dwarf followed by the bright spot where
the gas stream joins the accretion disc. We determine the system
parameters via a parameterized model of the eclipse fitted to the
observed lightcurve. We obtain a mass
ratio of $q = 0.570 \pm 0.011$ and an orbital inclination of
$84.6 \pm 0.3^{\circ}$. The primary mass is
$M_w = 0.69 \pm 0.07\,M_{\odot}$. The donor star's mass and radius
are found to be $M_d = 0.39 \pm 0.04\,M_{\odot}$ and
$R_d = 0.43 \pm 0.01\,R_{\odot}$, respectively. From the fluxes of
the white dwarf eclipse we find a white dwarf temperature of
$T_w = 23000 \pm 3000$~K, and a photometric distance to the system of
$1600 \pm 200$\,pc, neglecting the effects of interstellar reddening. The white dwarf temperature in KISJ1927 implies the white dwarf is accreting at an average rate of $\dot{M} = 1.4\pm0.8 \times10^{-9}$ M$_{\odot}$\,yr$^{-1}$, in
agreement with estimates of the secular mass loss rate from the donor. 
\end{abstract}


\bibliographystyle{mn2e_fixed2}

\section{Introduction} 
Cataclysmic variable stars (CVs) are a class of interacting binary system undergoing mass transfer, usually via a gas stream and accretion disc, from a Roche-lobe filling secondary to a white dwarf primary. In addition a `bright spot' is formed at the intersection of the disc and gas stream. \citet{warner95a} gives a comprehensive review of CVs. 

In recent years, continuous monitoring of CVs in the Kepler field have provided significant improvements in our understanding of accretion physics in close binaries. These observations have shed light on the strength of the disc viscosity \citep{o_Still_Howell_Wood_Smale_2010}, the origin of superhumps \citep{ill_Howell_Cannizzo_Smale_2011} and superoutbursts \citep{2013PASJ...65...50O} in the dwarf nova subclass of CVs, which exhibit quasi-periodic outbursts of a few magnitudes that arise in the accretion disc. 

Eclipsing dwarf novae in the Kepler field are particularly interesting, since modelling the eclipses allows the evolution of the accretion disc radius to be studied over the outburst cycle. The lightcurves of eclipsing dwarf novae can be quite complex, with the accretion disc, white dwarf and bright spot all being eclipsed in rapid succession. Foreshortening of the bright spot gives rise to an `orbital hump' in the lightcurve at phases $0.6-1.0$. With sufficient time-resolution, the eclipse structure allows the system parameters to be determined to a high degree of precision \citep{wood86a}. In turn, knowledge of the system parameters allows for more accurate modelling of the accretion disc dynamics. Eclipsing systems also allow us to measure the precise flux contribution of the white dwarf. With colours one can then estimate its temperature \cite[e.g][]{savoury11,littlefair06}. The temperatures of accreting white dwarfs reflect the accretion rate averaged over the past  $10^3$ to $10^5$ years \citep{Townsley_Gansicke_2009} and hence give a more representative picture of the long-term behaviour of the system than any present-day  measurement can achieve. This method requires that the contribution of the white dwarf is significant. For short period systems ($P < 2\,$h) this is typically the case, but the higher accretion rates of long period systems can often mask the white dwarf. Long period systems where the white dwarfs can be seen are hence of particular interest.

KIS J192748.53+444724.5 (hereafter KISJ1927) is the second eclipsing CV known in the Kepler field \citep{scaringi2013}, after V477 Lyr \citep{2012MNRAS.425.1479R}. It is a dwarf nova with an orbital period of 3.97 hours. Analysis of long-cadence (30 minute) Kepler data from a single quarter has already revealed disc radius variations over an outburst \citep{scaringi2013}. The Kepler lightcurves also show a strong bright spot and eclipses of the disc and white dwarf, although they lack the time resolution necessary to resolve them. Here we present high-time resolution ULTRACAM $u'g'r'$ lightcurves of KISJ1927 and use these lightcurves to  estimate  the system parameters.  The observations are described in section~\ref{sec:observations}, the results are presented in section~\ref{sec:results}, and discussed in section~\ref{sec:discussion}.

\section{Observations}
\label{sec:observations}
\begin{table*}
\begin{center}
\begin{tabular}{ccccccccc}
\hline
Date & Start Phase & End Phase  & $T_{mid}$  & Exp. time & NBLUE &
Data points  & Seeing  & Airmass \\ 
&&& BMJD(TDB)& seconds &  & & arcseconds  &  \\ \hline
2013 Aug 04 & 1834.89   & 1835.19   & 56508.92767(2) & 1.981 & 4 & 2156 & 1.2--1.4 & 1.25--1.05 \\
2013 Aug 04 & 1835.86   & 1836.24   & 56509.09294(2) & 1.981 & 4 & 2471 & 1.4--1.6 & 1.10--1.26 \\
2013 Aug 05 & 1840.74   & 1841.21   & 56509.91955(2) & 1.981 & 4 & 3392 & 1.2--1.3 & 1.15--1.05 \\
2013 Aug 05 & 1841.84   & 1842.14   & 56510.08486(2) & 1.981 & 4 & 2162 & 1.4--1.6 & 1.12--1.38 \\
\hline
\end{tabular}
\caption{\label{table:obs}Journal of observations. Most observing nights were clear of cloud, but the night of 4$^{th}$ Aug 2013 suffered from some light cirrus towards the end of observations. The dead-time between exposures was 0.025~s for all observations. The relative GPS timestamping on each data point is accurate to 50 $\mu$s, with an absolute accuracy better than 1 ms. $T_{mid}$ gives the time of mid eclipse (see section~\ref{subsec:times}). NBLUE indicates the number of blue frames which were co-added on-chip to reduce the impact of read noise.}
\end{center}
\end{table*}

On the nights of Aug 4$^{th}$ and Aug 5$^{th}$ 2013 KISJ1927 was observed simultaneously in the SDSS-$u'g'r'$ colour bands using ULTRACAM \citep{2007MNRAS.378..825D} on the 4.2-m William Herschel Telescope (WHT) on La Palma. A complete journal of observations is shown in table~\ref{table:obs}. Four eclipses were observed in total. Data reduction was carried out in a standard manner using the {\sc ultracam} pipeline reduction software, as described in \citet{feline04a}, and a nearby comparison star (separation 55\arcsec) was used to correct the data for transparency variations.  Using a second star to check for variability in the comparison star revealed its flux to be constant to better than 1 per cent. The photometry was transformed into the $u'g'r'$ standard system \citep{smith02} using observations of the standard star BD +33 2642 taken at the start of the night of  4$^{th}$ Aug. The photometry was corrected for extinction using nightly measurements of the
$r'$-band extinction from the Carlsberg Meridian Telescope (\url{http://www.ast.cam.ac.uk/~dwe/SRF/camc_extinction.html}), which was converted to an extinction in the $u'$ and $g'$ bands using the information provided in La Palma Technical Note 31 (\url{http://www.ing.iac.es/Astronomy/observing/manuals/ps/tech_notes/tn031.pdf}).

\section{Results}
\label{sec:results}

\subsection{Lightcurve morphology and variations}
\label{subsec:lcurves}
\begin{figure}
\begin{center}
\includegraphics[width=1.1\columnwidth]{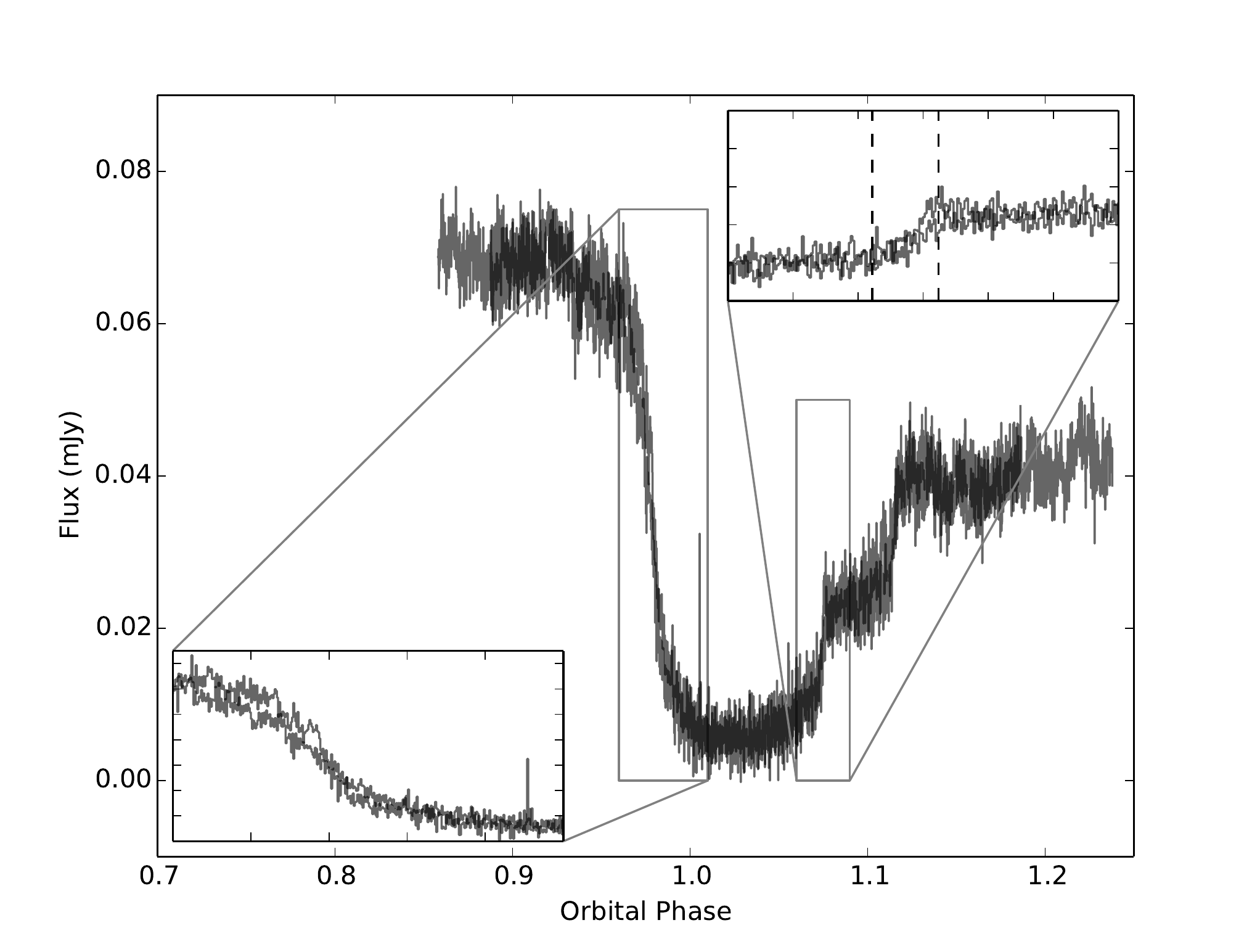}
\caption{\label{fig:lcurves04} $g'$-lightcurves of two eclipses from the night of Aug $4^{th}$ 2013, folded on the ephemeris of \protect\citet{scaringi2013}. Insets show the sections of the lightcurve around white dwarf/bright spot ingress and white dwarf egress. Dashed lines in the inset mark the beginning and end of the white dwarf egress, as determined from our best-fit model.}
\end{center}
\end{figure}

\begin{figure}
\begin{center}
\includegraphics[width=1.1\columnwidth]{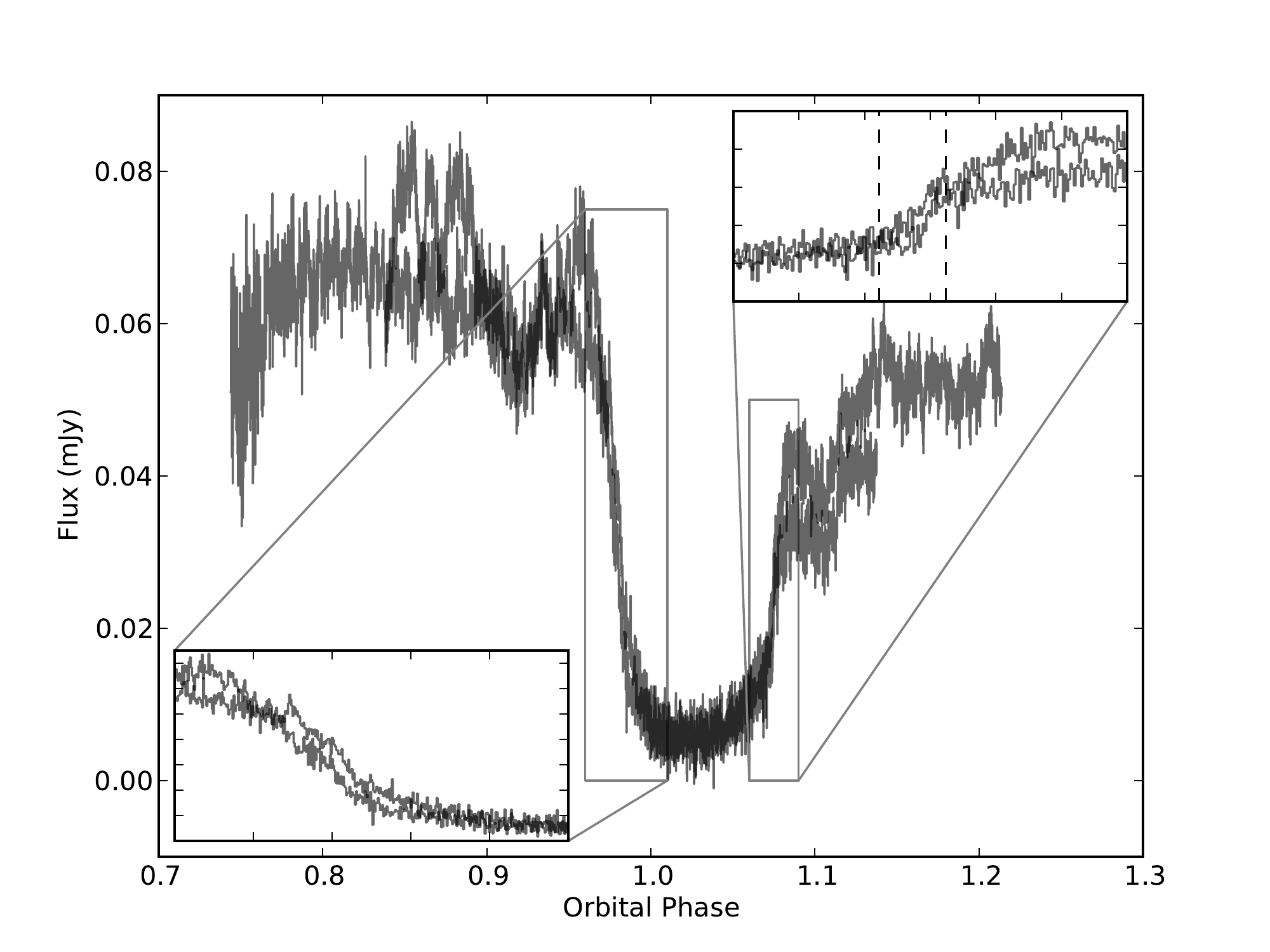}
\caption{\label{fig:lcurves05} $g'$-lightcurves of two eclipses from the night of Aug $5^{th}$ 2013, folded on the ephemeris of \protect\citet{scaringi2013}. Insets show the sections of the lightcurve around white dwarf/bright spot ingress and white dwarf egress. Dashed lines in the egress inset mark the beginning and end of the white dwarf egress, as determined from our best-fit model.}
\end{center}
\end{figure}

\begin{figure}
\begin{center}
\includegraphics[width=1.1\columnwidth,trim=100 60 0 0]{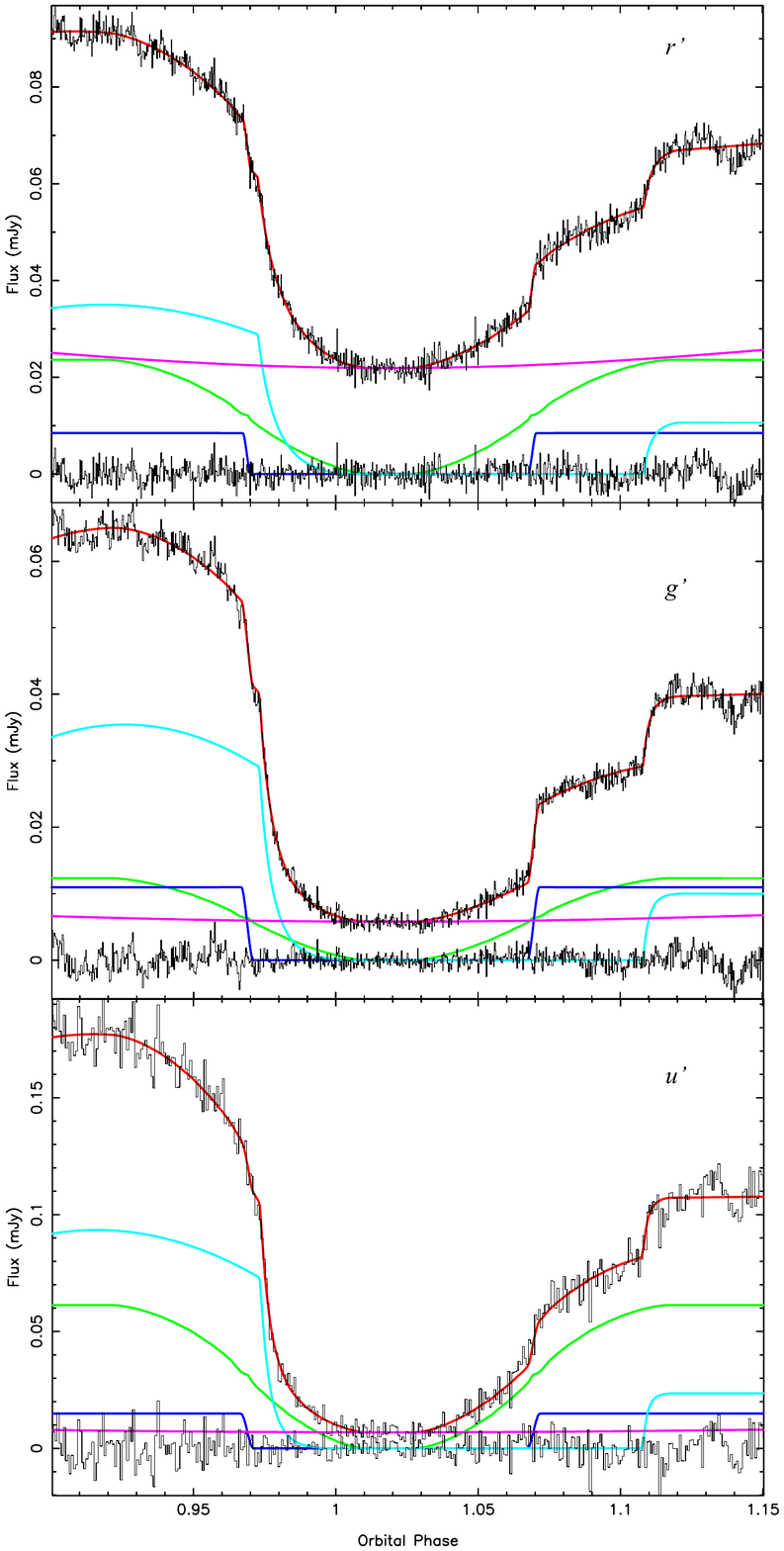}
\caption{\label{fig:fit}The phased-folded $u{{'}}g{{'}}r{{'}}$ light curves of KIS 1927, fitted using the model outlined in section~\ref{subsec:model}. In each panel the data (black) are shown with the fit (red) overlaid and the residuals plotted below (black). Also plotted are the separate light curves of the white dwarf (dark blue), bright-spot (light blue), accretion disc (green) and the secondary star (purple).}
\end{center}
\end{figure}

The individual eclipse lightcurves from the nights of Aug $4^{th}$ 2013 and Aug $5^{th}$ 2013 are shown in figures~\ref{fig:lcurves04}~and~\ref{fig:lcurves05} respectively. All data in this paper are shown folded on the ephemeris of \citet{scaringi2013}. The WHT/ULTRACAM data are taken at much higher time resolution than the long-cadence Kepler data of \citet{scaringi2013} (2 vs 1800 seconds) and reveal the detailed eclipse shape. The eclipses on the night of Aug $4^{th}$ are particularly clean, with little flickering, and show clear and separated egresses of the white dwarf and bright spot. The ingresses of the white dwarf and bright spot are harder to separate without averaging multiple eclipses. On the night of Aug $5^{th}$, the amplitude of flickering increased significantly,  being particularly high in the data taken around cycle 1841. Inspection of the region around white dwarf egress (see the inset of figure~\ref{fig:lcurves05}) shows that large amplitude flickering begins immediately following white dwarf egress, implying a physical location close to the white dwarf, either the inner disc or a boundary layer.  Due to the large amplitude flickering and extended bright spot ingress (which is likely to blend with any  white dwarf ingress features) we exclude cycle 1841 from the following determination of system parameters. Figure~\ref{fig:fit} shows that, in the phase-binned lightcurve excluding cycle 1841, the white dwarf and bright spot ingress are resolved, allowing a determination of the system parameters for KISJ1927.

\subsection{A parameterised model of the eclipse}
\label{subsec:model}
To determine the system parameters we used a physical model of the binary system to calculate eclipse lightcurves  similar to that first developed by \citet{e_Marsh_Cheng_Hubeny_Lanz_1994}. \citet{feline04b} showed that this method gives a more robust determination of the system parameters in the presence of flickering than the derivative method of \citet{wood86a}. Extracting  the system parameters from the eclipse lightcurve depends on four assumptions: that the bright spot lies on the ballistic trajectory from the secondary star; that the white dwarf is accurately described by a theoretical mass-radius relation; that the whole of the white dwarf is visible out of eclipse and that the donor star fills its Roche Lobe. We cannot be sure that all of these assumptions hold for every system, but we point out here that dynamical mass determinations of CVs agree with those found via the photometric method \citep{savoury12,copperwheat10,hillon_Littlefair_Savoury_2012,och_Rodriguez-Gil_Dhillon_2009}.

The model we fit is described in detail by \citet{savoury11}. To summarise, the free parameters of the model are:
\begin{enumerate}
\item the mass ratio, $q$;
\item the white dwarf eclipse phase full-width at half depth, $\Delta\phi$;
\item the (linear) white dwarf limb-darkening parameter, $U_w$;
\item the white dwarf radius, scaled to the orbital separation, $R_w/a$;
\item seven parameters describing the emission from the bright-spot;
\item a disc exponent, $b$, describing the power law of the radial intensity distribution of the disc;
\item a phase offset, $\phi_0$, and
\item the flux contributions of the white dwarf, donor star, accretion disc and bright spot. 
\end{enumerate}
The data are not of sufficient quality to constrain the white dwarf limb darkening parameter. Instead we perform an initial fit of the model described above to estimate the flux from the white dwarf in the $u'g'r'$ bands and to estimate the white dwarf temperature (see section~\ref{subsec:wdtemp}). Limb-darkening parameters for the appropriate white dwarf temperature and log $g$ are then interpolated from the tables in \citet{Strickland_Kilic_Bergeron_2013}. We adopt limb-darkening parameters of 0.516, 0.412 and 0.361 for $u'$, $g'$ and $r'$ respectively. Uncertainties on the limb darkening parameters lead to uncertainty in $R_w/a$ of $\sim1$\,per cent, which is small compared to our errors \citep{h_Gansicke_Baraffe_Watson_2007}. 

All parameters bar $U_w$ are then fitted to the lightcurve using the Markov-Chain Monte-Carlo method. The aim is to obtain a set of models whose parameters are distributed according to the posterior probability distribution of the model $M$, given the data $D$, i.e. $P(M|D)$. This is accomplished by stochastic jumping of the model parameters followed by selection or rejection according to $P(M) P(D|M)$, where $P(M)$ is the prior probability of the model, and $P(D|M)$ is the likelihood, which we estimate using $\exp (-\chi^2/2)$. The choice of distribution for jumping between models can drastically affect the efficiency and convergence of the MCMC fitting. During the initial, burn-in phase, we draw candidate models from a distribution defined by a covariance matrix which we obtain from an initial Levenberg-Marquardt fit. In the second phase we draw models from the distribution defined by the last 10,000 steps of the burn-in phase. These jumps are scaled by a factor which is chosen to give an acceptance rate of approximately 0.25, and fixed thereafter. We adopt uniform priors for most parameters. The exceptions are the bright spot, which is constrained to lie along a line within 40$^{\circ}$ of the tangent to the disc edge at the stream impact point, and a joint prior on $q$ and $\Delta\phi$, which forbids combinations which would imply an inclination greater than 90$^{\circ}$.

The $g'$- and $r'$-models were fit to the respective lightcurves using the above procedure. MCMC chains of 200,000 jumps were obtained. The first 100,000 jumps (the burn-in phase) are discarded and the second 100,000 steps used to estimate $P(M|D)$. We visually examined the chains for convergence, requiring that the means and root-mean-square values for all parameters showed no long term trends. The $u'$-data is not of sufficiently high quality to constrain the full model, but the $u'$-fluxes of the white dwarf are necessary to constrain the white dwarf temperature. We determine the relative contributions of the binary components to the $u'$-band light by holding the parameter set ($q$, $\Delta\phi$, $R_w/a$, $\phi_0$) fixed at the best-fit values from the $g'$-model and fitting the remaining parameters (bar $U_w$) via Levenberg-Marquardt minimisation. The results of this model fitting are only used to determine the $u'$-band fluxes of the various binary components. The model fits and phase-binned lightcurves are shown in figure~\ref{fig:fit}.

\subsection{White Dwarf Temperature and photometric distance}
\label{subsec:wdtemp}
We fit the $u'g'r'$ white dwarf fluxes with predictions from white dwarf model atmospheres \citep{rgeron_Wesemael_Beauchamp_1995} to derive the white dwarf temperature and distance to KISJ1927. Because the formal errors on our white dwarf fluxes do not take account of any uncertainties in our absolute photometry, we add systematic errors of 1 percent to the white dwarf fluxes. The $u'g'r'$ fluxes are fit simultaneously to the tabulated values in \cite{rgeron_Wesemael_Beauchamp_1995}, which depend upon log $g$ and effective temperature, with the distance to the white dwarf and the reddening as additional parameters. Parameters are optimised using MCMC, with uninformative priors on the effective temperature and distance. We initially set the priors to enforce no reddening.

An initial guess of the white dwarf temperature is derived by allowing log $g$ to vary as a free parameter. This guess is then used in the lightcurve fitting to derive log $g$, as described in section~\ref{subsec:params}. Our final white dwarf temperature and distance are then derived with a prior on log $g$ which enforces the constraints suggested by the light curve fitting, including the  uncertainties in log $g$. The suggested white dwarf temperature is $T_{wd} = 23000 \pm 3000$\ K. The distance is found to be $d=1600 \pm 200$\ pc, where the error is dominated by the uncertainties in log g. These formal errors do not include any systematic effects, including contamination of our white dwarf colours, or the effect of neglecting reddening. 

\begin{figure}
\begin{center}
\includegraphics[width=1.1\columnwidth,trim=50 50 50 50]{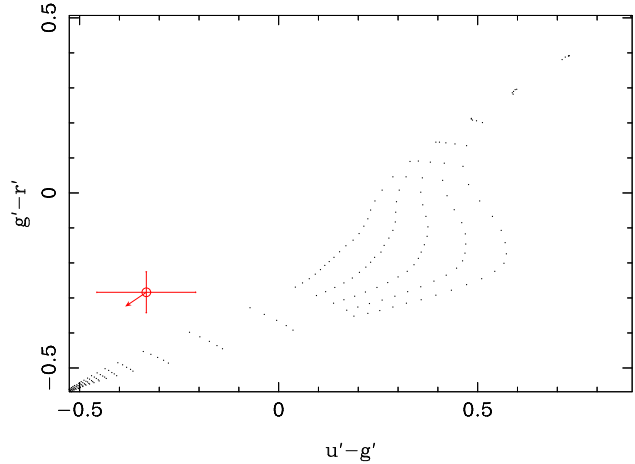}
\caption{\label{fig:wdcolours} The white dwarf colours derived from our model plotted together with the white dwarf models of \protect\citet{rgeron_Wesemael_Beauchamp_1995}. From top to bottom, each curve represents $\log g = 9.0, 8.5, 8.0, 7.5$ and $7.0$, respectively. The measured white dwarf colours are shown here in red, and are used to derive the white dwarf temperature, which in turn is used to correct the white dwarf mass{\textendash}radius relationships used later to obtain the final system parameters. An arrow marks the shift in observed white dwarf colours if they were de-reddened assuming a Galactic reddening of $E(B-V)=0.121$. }
\end{center}
\end{figure}

Figure~\ref{fig:wdcolours} shows that the white dwarf colours lie about 2$\sigma$ outside the range predicted by the model atmospheres. It is therefore likely that our white dwarf colours are affected by contamination from the disc or bright-spot, or an un-modelled light source such as the boundary layer. Another possibility is that the flickering which begins immediately after the egress of the white dwarf is affecting our estimate of the white dwarf colours. If our white dwarf colours are incorrect, then our derived white dwarf temperature and photometric distance will be affected.  An additional source of uncertainty is the extinction towards KISJ1927, which is unknown. The Galactic extinction in this direction is $E(B-V)=0.121$ \citep{schlafly11}, and with a distance in excess of a kilo-parsec a substantial amount of this may be in front of the CV. If we relax the prior constraint on the reddening to be between zero and $E(B-V)=0.121$, we obtain measurements of the white dwarf temperature and distance of $T_{wd} = 28000 \pm 7000$\ K and $d=1700 \pm 400$\ pc.  The posterior probability distribution for the reddening does not allow us to rule out any values less than $E(B-V)=0.121$, but the most likely value is $E(B-V)=0.09$.

We can use the colours of the donor star as an independent check on the reddening towards KISJ1927. We estimate the donor flux in each band using the results of the model fitting described in section~\ref{subsec:model}. This implies donor magnitudes of $r'=20.17\pm0.01$ and $g'=21.64\pm0.01$. The $g'-r'$ colour of $1.47\pm0.02$ is in reasonable agreement with the spectral type of M3 expected for a donor star at this orbital period \citep{Knigge_Baraffe_Patterson_2011,Covey2007}. If the reddening towards KISJ1927 is as high as $E(B-V)=0.121$, it would imply an intrinsic $g'-r'$ colour for the donor star of $\sim1.33$.  In turn this would suggest a spectral type of roughly M0, or a mass of $\sim 0.6$\,M$_{\odot}$. This is much earlier than the evolutionary models of \cite{Knigge_Baraffe_Patterson_2011} predict for a CV of this orbital period, and the mass is higher than the donor mass measured in section~\ref{subsec:params}. For this reason we adopt the white dwarf parameters assuming low reddening towards KISJ1927. However, we must point out that  $g'-r'$ is almost constant for spectral types later than M0, \citep{finlator10}, and we  cannot rule out higher reddening values.

The donor star colours can also provide an independent constraint on the distance. The $g'-r'$ colour can be used to estimate the surface brightness which, in combination with the apparent magnitude of the donor, constrains the distance. The zero-magnitude angular diameter, $\theta_{(m_{\lambda}=0)}$, is a measure of surface brightness, defined as the angular diameter a star would have if its apparent magnitude $m_{\lambda} = 0$. Thus, if $\theta$ is the actual angular diameter of the star, 
\begin{equation}
\log_{10}  \theta_{(m_{\lambda}=0)} = \log_{10} \theta + 0.2 m_{\lambda}.
\label{eq:size}
\end{equation}
The distance to the star is therefore given by
\begin{equation}
\log_{10} d = \log_{10} (2R) + 0.2 m_{\lambda} - \log_{10}  \theta_{(m_{\lambda}=0)},
\label{eq:dist}
\end{equation}
where $d$ is measured in parsecs, $R$ is the radius of the star, measured in AU and the angular diameter is measured in arc seconds. \cite{boyajian13} have tabulated $\theta_{(g'=0)}$ as a function of $g'-r'$ colour using a compilation of interferometric observations of stars. Using equation~\ref{eq:size} we find
\begin{equation*}
\log_{10}  \theta_{(r'=0)}  = \log_{10}  \theta_{(g'=0)}  - 0.2 (g'-r')
\end{equation*}
which we use in combination with equation 5 of \cite{boyajian13}, assuming Solar metallicity, to find $\log_{10}  \theta_{(r'=0)}  = -1.53 \pm 0.05$. Using equation~\ref{eq:dist} and our $r'$-band donor magnitude we find $d = 1500 \pm 200$\ pc, in good agreement with the white dwarf photometric distance.

\subsection{System Parameters}
\label{subsec:params}
\begin{figure*}
\begin{center}
\includegraphics[width=2.0\columnwidth,trim=50 50 50 50]{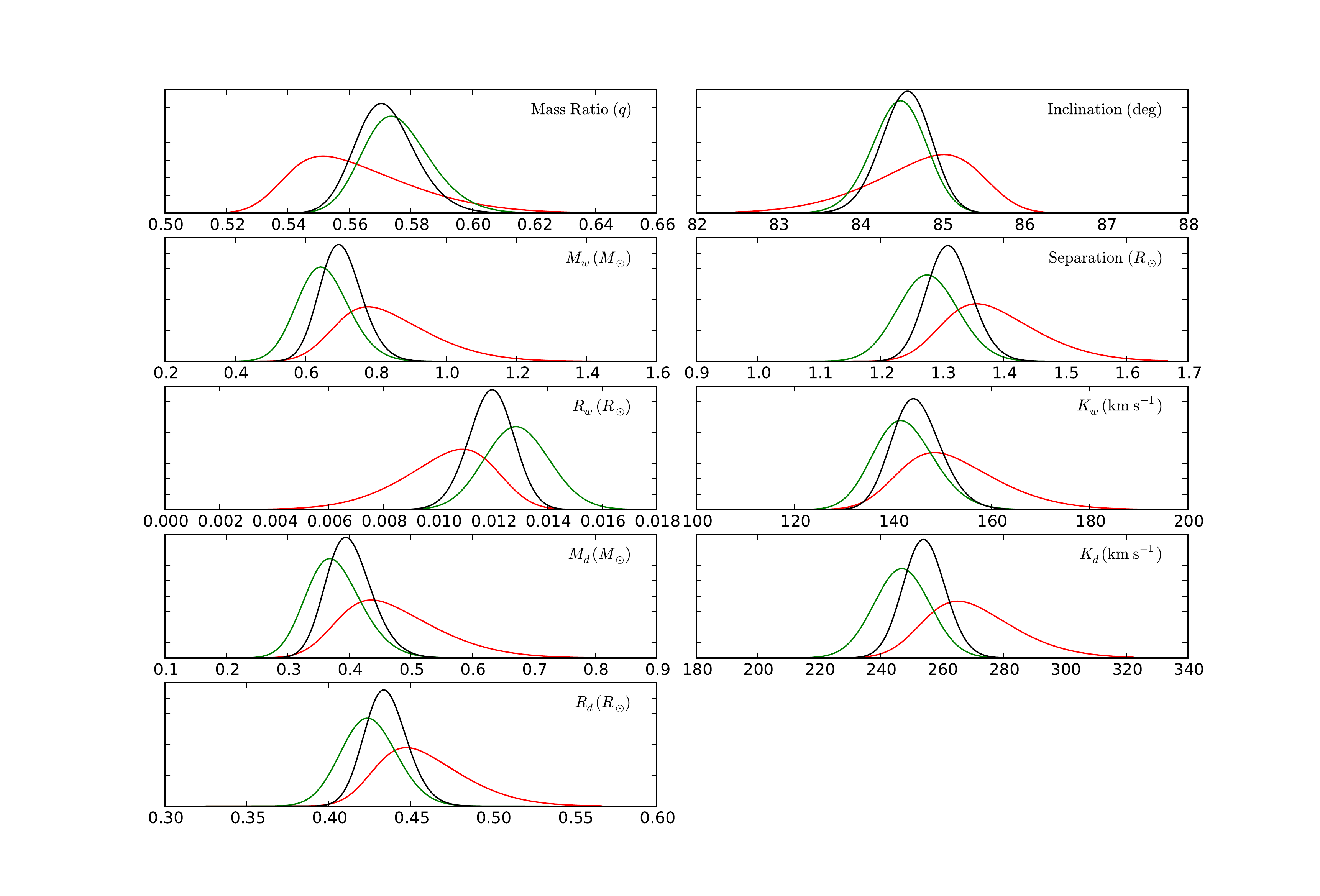}
\caption{\label{fig:params} The normalised posterior probability density functions (PDFs) for the system parameters of KISJ1927. Each PDF was derived using the MCMC chains for model fit parameters described in section~\protect\ref{subsec:model}, combined with Kepler's third law and a mass-radius relationship for the white dwarf component corrected to the appropriate effective temperature.The red curve represents the PDF from fitting the $r'$-band lightcurve, the green curve the PDF from the $g'$-band lightcurve and the black represents the total, combined, PDF. PDFs are shown for the mass ratio, $q$, the white dwarf mass and radius $M_w, R_w$, the donor mass and radius $M_d, R_d$, the inclination $i$, orbital separation $a$ and the radial velocities of the white dwarf ($K_w$) and donor ($K_d$). }
\end{center}
\end{figure*}

Our MCMC chains provide an estimate of the joint posterior probability distributions of $(q, \Delta\phi, R_w/a)$. At each step in the chain we can calculate the system parameters from these three values, Kepler{'}s third law, the orbital period, and a white dwarf mass{\textendash}radius relationship (corrected to our derived white dwarf temperature). We favour the mass-radius relationships of \citet{Wood_1995}, because they have thicker hydrogen layers which may be more appropriate for CVs. Adopting other white dwarf models \cite[e.g][]{panei00} changes our results by less than 1 per cent. The result is an estimate of the posterior probability distribution for the mass ratio $q$, white dwarf mass $M_w$, white dwarf radius, $R_w$, white dwarf gravity $\log g$, donor mass $M_d$, donor radius $R_d$, inclination $i$, binary separation $a$, and the radial velocities of the white dwarf and donor star ($K_w$ and $K_d$, respectively). We can then combine the posterior distributions obtained for each band into the total posterior distribution for the system parameters, as shown in figure~\ref{fig:params}. We note that most parameters show a Gaussian distribution with very little asymmetry. Our adopted value for a given parameter is taken from the peak of the posterior distribution function. Upper and lower error bounds are derived from the 67 per cent confidence levels. 

Since the white dwarf temperature is used to correct the adopted white dwarf mass-radius relationship, uncertainties in white dwarf temperature will have a knock-on effect on the derived system parameters. \citet{savoury11} examined the consequences of a 2000\,K systematic error in the white dwarf temperature, and found it introduced 
negligible errors into the white dwarf mass and radius, and distance errors of $\sim20$\ pc. Therefore, we do not believe that uncertainty in the white dwarf temperature is significantly affecting our derived system parameters.
\begin{table}
\begin{center}
\def\arraystretch{1.5}
\begin{tabular}{l|ccc}
\hline
& $r'$ & $g'$ & Combined  \\ 
\hline
$q$  &  $0.551^{+0.03}_{-0.01}$ & $0.574^{+0.013}_{-0.008}$ & $0.570\pm0.011$\\ 
$M_w\ (M_{\odot})$  &  $0.78^{+0.2}_{-0.1}$ & $0.64^{+0.08}_{-0.06}$ & $0.69\pm0.07$\\ 
$R_w\ (R_{\odot})$  &  $0.011^{+0.001}_{-0.002}$ & $0.013^{+0.001}_{-0.001}$ & $0.0120\pm0.0009$\\
${\rm log\,}g$  &  $8.3^{+0.3}_{-0.2}$ & $8.0^{+0.1}_{-0.1}$ & $8.1\pm0.3$\\ 
$T_w\ ({\rm K})$  &  &  & $23000\pm3000$\\ 
$M_d\ (M_{\odot})$  &  $0.44^{+0.12}_{-0.04}$ & $0.37^{+0.05}_{-0.04}$ & $0.39\pm0.04$\\ 
$R_d\ (R_{\odot})$  &  $0.45^{+0.04}_{-0.02}$ & $0.42^{+0.02}_{-0.02}$ & $0.43\pm0.01$\\
$a\ (R_{\odot})$  &  $1.36^{+0.11}_{-0.05}$ & $1.28^{+0.05}_{-0.05}$ & $1.31\pm0.04$\\ 
$K_w\ ({\rm km\,s}^{-1})$  &  $149^{+13}_{-7}$ & $141^{+7}_{-5}$ & $144\pm6$\\
$K_r\ ({\rm km\,s}^{-1})$  &  $266^{+21}_{-11}$ & $247^{+9}_{-9}$ & $254\pm7$\\ 
$i\ (^{\circ})$  &  $85.0^{+0.4}_{-1.0}$ & $84.5^{+0.3}_{-0.4}$ & $84.6\pm0.3$\\ 
$d\ ({\rm pc})$  &   &  & $1600\pm200$\\
\hline
\end{tabular}
\caption{\label{tab:params}System parameters for KISJ1927 as derived from the PDFs shown in figure~\protect\ref{fig:params}. $R_d$ is the volume radius of the donor star's Roche lobe as defined by \protect\citet{Eggleton_1983}.  }
\end{center}
\end{table}

The resulting system parameters are listed in table~\ref{tab:params}. Note the uncertainties in table~\ref{tab:params} reflect the uncertainties from our MCMC fitting alone; they do not account for uncertainties arising from the failure of the assumptions listed in section~\ref{subsec:model}, nor do they account for the effects of flickering. Flickering can be considered an additional noise source which we do not model; perhaps the best way to reduce the impact of flickering is to average many eclipses. Unfortunately, we only have three suitable eclipses of this system to combine. As a result, continued monitoring of KISJ1927 is needed to improve both the accuracy and the precision of the system parameters.

\subsection{Eclipse Times}
\label{subsec:times}
Mid-eclipse times for each eclipse were determined by fitting the model described in section~\ref{subsec:model} to each eclipse individually. Parameters which ought to be time-invariant $(q, \Delta\phi, U_w, R_w/a)$ were held fixed at the best fit values derived from the average $g'$ lightcurve, whilst all other parameters, including the phase offset $\phi_0$ were allowed to vary. An initial fit was performed using an implementation of the  simplex algorithm. A final value for $\phi_0$, and corresponding uncertainty, was found using a further fitting step using a Levenberg-Marquardt algorithm to minimise $\chi^2$. Mid-eclipse times are shown in table~\ref{table:obs}.
We note here that our eclipse times are for the middle of white dwarf eclipse, whilst the ephemeris in \citet{scaringi2013} gives the times of minimum light, after smoothing by a 1800-second running mean. Comparing our mid-eclipse times with the ephemeris 
in \citet{scaringi2013} suggests that, for our observations, the predicted time of minimum light is offset from the white dwarf mid-eclipse by 0.0031(2) days, as one would expect given the overall asymmetry of the eclipse.

\section{Discussion}
\label{sec:discussion}
The presence of both stable (nova-likes) and unstable (dwarf novae) accreting CVs at the same orbital period is a long-standing puzzle. According to the disc instability model (DIM), the difference between nova-likes and dwarf novae is the rate at which mass is supplied to the disk. Above some critical rate, the disk is bright and stable, and the system is a nova-like. At lower rates, the system shows dwarf nova outbursts. At an orbital period of $\sim3.97$ hours, KISJ1927 is (like most systems a period of 4 hours) a dwarf nova \cite[see][for example]{Knigge_Baraffe_Patterson_2011}. Mass transfer rates in CVs are driven by angular momentum loss. For long period CVs ($P > 3$\ hours), the angular momentum loss is thought to be dominated by magnetic wind braking, whilst the shorter period CVs have much lower rates of wind braking, and angular momentum losses dominated by gravitational radiation. By tuning the angular momentum loss rates to fit the location of CV donors in a mass-orbital period diagram, \cite{Knigge_Baraffe_Patterson_2011} obtain a semi-empirical estimate of the secular mass transfer rates in CVs. Somewhat surprisingly, the secular mass transfer rate is everywhere {\em lower} than the critical rate required for stable accretion, and especially so in the period range 3--4 hours, where the novalike variables are strongly concentrated. Viewed in this light, the observed existence of nova-likes at any orbital period can be viewed as surprising. 

There are several ways to explain the existence of nova-like variables. It is possible that short-term cycles in mass transfer rates exists. Nova-likes would then be systems in the high state of a cycle. However, it is also possible that estimates of the critical mass transfer rate \citep[e.g.][]{2001NewAR..45..449L}, or the semi-empirical estimate of the secular mass transfer rate \citep{Knigge_Baraffe_Patterson_2011} are in error. 

Temperature estimates for white dwarfs with well-known masses can help address these questions. Together, the white dwarf temperature and mass provide a measure of the average accretion rate over a timescale of order $10^3$--$10^5$ years \citep{Townsley_Bildsten_2003}. This timescale is too short to provide a reliable estimate of the {\em secular} mass transfer rate \citep{Townsley_Gansicke_2009}, and so the mass transfer rate from white dwarf temperatures can be compared to estimates of the secular rate to look for evidence of mass-transfer rate cycles.

With an orbital period of $\sim3.97$ hours, KISJ1927 is similar to the well studied system IP Peg, which has an orbital period of $\sim 3.8$ hours.  The white dwarf temperature and mass in KISJ1927 implies an average mass transfer rate over the medium term of $\dot{M} = 1.4 \pm 0.8 \times 10^{-9}\,{\rm M}_{\odot} {\rm yr}^{-1}$, which compares well with estimates of the secular mass transfer rate at this orbital period  of  $\dot{M} \sim 2 \times 10^{-9}\,{\rm M}_{\odot} {\rm yr}^{-1}$ \citep{Knigge_Baraffe_Patterson_2011}. The agreement between these rates suggests that the critical mass transfer rate for stability exceeds the secular rate at these periods, and that uncertainties in these rates may not explain the dominance of novalike variables between 3--4 hours. 

By contrast to KISJ1927, IP Peg's white dwarf temperature is much cooler, despite the higher mass white dwarf, implying a mass transfer rate of $\dot{M} <5\times10^{-11} {\rm M}_{\odot} {\rm yr}^{-1}$ \citep{ckman_Gansicke_Southworth_2010}. The low temperature in IP Peg is well below the expected secular mass transfer rate. This may be evidence for mass transfer rate cycles which long enough to alter the white dwarf temperature, or perhaps that IP Peg has only begun mass transfer recently, and the white dwarf has yet to reach equilibrium temperature \citep{ckman_Gansicke_Southworth_2010,otovic_Schreiber_Gansicke_2011}. 

A final possibility is that the white dwarf temperature in KISJ1927 is higher than that in IP Peg due to heating during  a recent dwarf nova outburst \citep[e.g][]{godon2003}. This might affect measurements of the white dwarf temperature for a duration of a few days to weeks after the outburst. The public Kepler data for KISJ1927 shows three outbursts, with quiescent periods lasting 94 and 59 days between them. The most recent outburst in the Kepler data is 143 days prior to our observations. However, the Kepler data stops 88 days before our observations due to the reaction wheel failures on the Kepler spacecraft. A high white dwarf temperature due to an outburst remains a possibility.

We stress that the conclusions reached here regarding the white dwarf temperature in KISJ9172 are preliminary. The white dwarf temperature has been determined from the colours of white dwarf ingress/egress in only three eclipses. As discussed in section~\ref{subsec:wdtemp} it is possible that systematic errors affect our white dwarf temperature. Indeed, \citet{ckman_Gansicke_Southworth_2010} found that the colours of white dwarf egress in IP Peg were variable between eclipses, and attribute this to changing obscuration of the white dwarf by the inner disc. If, however, our preliminary estimate of the white dwarf temperature turns out to be accurate, the implication is that KISJ1927 has an average $\dot{M}$ in line with expectations of the secular rate. 

The component masses of KISJ1927 compare reasonably well with expectations. The white dwarf mass of $0.69\pm0.07 M_{\odot}$ is comparable to the average white dwarf mass in CVs of $0.79 \pm 0.03 M_{\odot}$ \citep{eat_Kerry_Hickman_Parsons_2011}. As a result, we can compare our donor star properties directly to the revised theoretical tracks of \citet{Knigge_Baraffe_Patterson_2011}, which are computed for a white dwarf mass of $0.75M_{\odot}$. There is mild evidence that the donor star in KISJ1927 is slightly more massive and larger than expected; the predicted donor mass at this period is $0.32M_{\odot}$, compared to a measured value of $0.39\pm0.04M_{\odot}$, and the predicted donor radius is $0.40R_{\odot}$, compared to our measured value of $0.43\pm0.01M_{\odot}$. It will be interesting to see if this discrepancy survives the test of additional data.

\section{Conclusions}
\label{sec:conclusions}
We present high-speed, three-colour photometry of KISJ1927, an eclipsing dwarf nova in the Kepler field. Taken over two successive nights, the eclipses show appreciable evolution of the flickering and hot-spot features over timescales as short as one orbit. Averaging of the three eclipses least affected by flickering allows us to separate the white dwarf and bright spot ingress and hence obtain a photometric measure of the system parameters. We obtain a mass ratio of $q = 0.570\pm0.011$ and an orbital inclination of $84.6 \pm 0.3^{\circ}$. The primary mass is $M_w = 0.69\pm0.07M_{\odot}$. The donor star's mass and radius are found to be $M_d = 0.39\pm0.04M{\odot}$ and $R_d = 0.43\pm0.01R{\odot}$, respectively. From the fluxes of the white dwarf eclipse we find a white dwarf temperature of $T_w = 23000 \pm 3000$\ K, and a photometric distance to the system of $1600 \pm 200$\,pc. These system parameters will improve the utility of KISJ1927 as a test bed for accretion disc physics using the continuous coverage of the Kepler satellite.

Whilst the system parameters of KISJ1927 are in line with theoretical expectations for a CV of this orbital period, the white dwarf temperature is strikingly different from that of IP Peg, a dwarf nova at a similar orbital period. This large difference in temperatures  may be explained either by the recent onset of mass transfer in IP Peg, or by the presence of relatively long-term mass transfer rate cycles in CVs at this orbital period.

\section{Acknowledgements}
We thank the anonymous referee for a careful reading of this text. VSD, SPL and ULTRACAM are supported by STFC grant ST/J001589/1. The results presented in this paper are based on observations made with the William Herschel Telescope operated on the island of La Palma by the Isaac Newton Group in the Spanish Observatorio del Roque de los Muchachos of the Instituto de Astrofisica de Canarias. This research has made use of NASA{'}s Astrophysics Data System Bibliographic
Services and the SIMBAD data base, operated at CDS, Strasbourg, France. 

\bibliography{refs,refs2}

\end{document}